\documentclass[10pt,a4paper]{article}
\usepackage{amsfonts}
\usepackage[english]{babel}
\usepackage[ansinew]{inputenc}
\usepackage{amsmath}
\usepackage{amssymb}
\usepackage[pdftex]{color}
\usepackage{lscape}
\usepackage{multirow}
\usepackage{arydshln}
\usepackage{url,floatflt}
\usepackage{subfig}
\usepackage{hyperref}
\usepackage[table]{xcolor}
\usepackage{tikz}
\usetikzlibrary{shapes,snakes}
\newcommand*\circled[1]{\tikz[baseline=(char.base)]{
            \node[shape=circle,draw,inner sep=2pt] (char) {#1};}}
            \newcommand*\squared[1]{\tikz[baseline=(char.base)]{
            \node[shape=rectangle,draw,inner sep=2pt] (char) {#1};}}
\newcommand*\diamonded[1]{\tikz[baseline=(char.base)]{
            \node[shape=diamond,draw,inner sep=2pt,scale=0.9] (char) {#1};}}

\newcommand{\Zset}{\mathbb{Z}}
\newcommand{\Fset}{\mathbb{F}}
\newcommand{\zech}[2]{\mathcal{Z}_{#1}({#2})}
\newcommand{\card}[1]{\left|#1\right|}

\usepackage{geometry}

\usepackage{cancel}

\RequirePackage{framed}
\RequirePackage[amsmath,thmmarks]{ntheorem}\theoremseparator{:}
\theorembodyfont{\upshape}
\newtheorem{example}{Example}

\newtheorem{proof}{Proof}
\newtheorem{definition}{Definition}
 \newtheorem{remark}{Remark}
 
 \newtheorem{proposition}{Proposition}
 \newtheorem{corollary}{Corollary}

\begin {document}
\title{Recovering decimation-based cryptographic sequences by means of linear CAs\footnote{The first author was partially support by S\~{a}o Paulo State Research Council (FAPESP) grant 2015/07246-0 and CAPES. This research has been partially supported by Ministerio de Econom\'ia, Industria y Competitividad (MINECO), Agencia Estatal de Investigaci\'on (AEI), and Fondo Europeo de Desarrollo Regional (FEDER, UE) under project COPCIS, reference TIN2017-84844-C2-1-R, and by Comunidad de Madrid (Spain) under project reference S2013/ICE-3095-CIBERDINE-CM, also co-funded by European Union FEDER funds.}}
\author{Sara D. Cardell$^1$ and Amparo F\'uster-Sabater$^2$\\
\small $^1$Imecc, Unicamp, Campinas, Brazil\\
\small$^2$ITEFI, CSIC, Madrid,  Spain
}
\date{}
\maketitle
\begin{abstract}
The sequences produced by the cryptographic sequence generator known as the shrinking generator can be modelled as the output sequences of linear elementary cellular automata. 
These sequences are composed of interleaved m-sequences produced by linear structures based on feedback shifts.
This profitable characteristic can be used in the cryptanalysis of this generator. 
In this work we propose an algorithm that takes advantage of the inherent linearity of these cellular automata and the interleaved m-sequences. 
Although irregularly decimated generators have been conceived and designed as non-linear sequence generators, in practice they can be easily analysed in terms of simple linear structures.
%

\end{abstract}
\begin{small}
\begin{tabular}{l}
\textbf{keywords}: \textit{decimation, shrinking generator, cellular automata, Zech logarithm, cryptanalysis} 
\end{tabular}
\end{small}
\section{Introduction}\label{sec:intro}

Stream ciphers are cryptographic primitives used to ensure privacy in digital communication~\cite{Robshaw2008bk}.
Their  procedure consists in generating a long sequence as random as possible,  called the keystream sequence, from a short secret key and a public algorithm (the sequence generator).
For encryption, the sender performs a bit-wise XOR operation among the bits of the  keystream sequence and the message (or plaintext).
The resultant message (or ciphertext) is sent to the receiver.
For decryption, the receiver generates the same keystream and performs the same bit-wise XOR operation between the ciphertext and the keystream to recover the original message.

Most keystream generators are based on maximal-length Linear Feedback Shift Registers (LFSRs)~\cite{Golomb1982bk}  whose output sequences, the so-called m-sequences, are combined by means of nonlinear functions to produce pseudorandom sequences for cryptographic applications. Desirable properties for such sequences can be enumerated as follows:
a) Long Period,
b) Good statistical properties,
c) Large Linear Complexity~\cite{Peinado2013}.

Among the current keystream generators used in this type of cryptography, the class of irregularly decimated generators is one of the most popular \cite{Batina2004, Paar2010bk}.
The underlying idea of this kind of generators is the irregular decimation of an m-sequence according to the bits of another m-sequence.
The result of this decimation process is a new sequence that will be used as keystream sequence in stream ciphers.
One of the most representative elements of this family is the shrinking generator introduced by Coppersmith \textit{et al}. in \cite{Coppersmith1993}.
The sequence produced by this generator is called the shrunken sequence and has good cryptographic properties.

In \cite{Cardell2016f}, the authors showed that the shrunken sequence is composed of interleaving m-sequences generated by the same primitive polynomial.
Furthermore, in \cite{Cardell2016f} it was proven that the shrunken sequence can be modelled as one of the (vertical) sequences produced by a linear cellular automata (CAs).
This CA produces other sequences, which we call companion sequences of the shrunken sequence.
Such sequences have the same characteristic polynomial as the shrunken sequence and are composed of interleaving m-sequences as well.
In this work, we use all these properties to propose an algorithm that recovers the shrunken sequence given a small quantity of intercepted bits of such a sequence. In fact, the number of bits needed for the cryptanalysis is dramatically reduced when compared with the shrunken sequence period.

The paper is organized as follows: In Section~\ref{sec:prel}, some basic concepts and definitions are provided for the understanding of the work.
In Section~\ref{sec:mod}, we introduce some properties of the shrunken sequence as well as the characterization of the 102-CA that generates the shrunken sequence as one of its (vertical) sequences.
Next, in Section~\ref{sec:zech}, we review the main properties of the Zech logarithm, which will play an important role in the recovering algorithm.
In Section~\ref{sec:cryp}, an algorithm to recover the initial state of  both registers  in the shrinking generator is proposed.
Finally, the paper concludes with Section~\ref{sec:con}.
\section{Preliminaries}\label{sec:prel}
In this section we introduce the basics to understand the next sections.
In Subsection~\ref{sec:basic}, we remind some basic concepts about sequences and cryptography.
In Subsections~\ref{sec:SG} and \ref{sec:CA}, we provide the definition of the main concepts used throughout this paper: shrinking generator and cellular automaton.

\subsection{Basic concepts}\label{sec:basic}
Let $\mathbb{F}_2$ be the Galois field of two elements.
A sequence $\{a_i\}=\{a_0, a_1, \ldots\}$ is a binary sequence or a sequence  over $\mathbb{F}_2$  if its terms $a_i\in \mathbb{F}_2$, for $i=0, 1, \ldots$.
Besides, the sequence $\{a_i\}$ is said to be \textbf{periodic} if and only if there exists an integer $T$ such that $a_{i+T}=a_i$, for all $i\geq 0$.

Let $l$ be a positive integer, and let $c_0, c_1, \ldots, c_{l-1}\in\Fset_2$.
A binary sequence $\{a_i\}$   satisfying the recurrence relation
\begin{equation}\label{eq:22}	
a_{i+l}=c_0a_i + c_1a_{i+1}+ \cdots +c_{l-2}a_{i+l-2}+c_{l-1}a_{i+l-1}, \quad i\geq 0,
\end{equation}
is called an ($l$-th order) \textbf{linear recurring sequence} in $\mathbb{F}_2$.
The  first $l$ terms $\{a_0, a_1, \ldots,$ $a_{l-1}\}$ determine the rest of the sequence uniquely and are known as the \textbf{initial state}.
A relation of the form given in expression (\ref{eq:22}) is called an ($l$-th order) linear recurrence relation.

The monic polynomial of degree $l$ given by
 $$p(x)=c_0 + c_1x+ \cdots +c_{l-2}x^{l-2}+c_{l-1}x^{l-1}+x^l\in \mathbb{F}_2[x],$$
is called the \textbf{characteristic polynomial} of the  sequence $\{a_i\}$ and this is said to be generated by $p(x)$.

The generation of linear recurring sequences can be implemented by \textbf{Linear Feedback Shift Registers} (LFSRs) \cite{Golomb1982bk}.
These  are electronic devices with $l$ memory cells or stages which handle information in the form of elements of $\mathbb{F}_2$ and that are based on shifts and linear feedback.
If the characteristic polynomial $p(x)$ is primitive, then the LFSR is said to be maximal-length and the output sequence has period $2^{l}-1$.
This output sequence is called  \textbf{m-sequence} (maximal sequence) or \textbf{PN-sequence }(pseudonoise sequence).
\begin{example}
Given the primitive polynomial $p(x)=1+x^2+x^5$, the linear recurrence relation is given by:
\[
a_{i+5}=a_{i+2}+a_{i}, \text{ for } i=0, 1, 2, \ldots
\]
Given a initial state $\{1, 0, 0, 0, 0\}$, the m-sequence generated is the following
\[
\{
   1 \  0\   0 \  0  \ 0  \ 1  \ 0  \ 0 \ 1  \ 0 \  1 \  1 \  0  \ 0 \  1\  1 \  1 \  1\   1 \  0 \   0  \ 0 \  1 \  1 \  0  \ 1 \  1 \  1  \ 0 \  1 \  0\}
\]
with maximum period $2^5 - 1 = 31$.\hfill $\blacksquare$
\end{example}

The \textbf{linear complexity}, $LC$, of a sequence $\{a_i\}$ is defined as the length of the shortest LFSR that  generates such a sequence.
The linear complexity is related with the amount of sequence needed to reconstruct the whole sequence.
In cryptographic terms, the linear complexity must be as large as possible in order to prevent the application of the Berlekamp-Massey algorithm \cite{Massey1969}.
This algorithm efficiently computes the length and characteristic polynomial of the shortest LFSR given at least $2LC$ sequence bits.
A recommended value for $LC$ is about half the sequence period.

Due to their low linear complexity, LFSRs are never used alone as keystream generators.
In fact, m-sequences generated by LFSRs have good statistical properties, desirable for keystream design, but their linearity has to be destroyed, i.e., their linear complexity has to be increased before such sequences are used for cryptographic purposes.

\subsection{The shrinking generator}\label{sec:SG}
The \textbf{shrinking generator} was first introduced by Coppersmith \emph{et al.} in \cite{Coppersmith1993}.
It consists of two m-sequences where one of them decimates the other one.
Given two m-sequences $\{a_i\}$ and $\{b_i\}$ $(i \geq 0)$ generated by two LFSRs of length $L_1$ and $L_2$, respectively, the decimation rule is very simple:
$$\begin{cases}
\text{If } a_i=1 \text{ then } s_j=b_i,\\
\text{If } a_i=0 \text{ then } b_i \text{ is discarded},
\end{cases}$$
where the generated sequence, $\{s_j\}$ $(j \geq 0)$, is said to be the \textbf{shrunken sequence}.
If $\gcd(L_1,L_2)=1$, then the period of such sequence is $T=(2^{L_2}-1)2^{L_1-1}$.
Furthermore, the characteristic polynomial of this sequence is given by $p(x)^{L}$, where $p(x)$ is a primitive polynomial of degree $L_2$ and $ 2^{L_1-2}< L \leq 2^{L_1-1}$.
Therefore, the linear complexity of the shrunken sequence is given by $LC=L\cdot L_2$.
As usual, the key of this generator is the initial state of both registers.

\begin{example}\label{ex:1}
Consider the register $R_1$ with characteristic polynomial $p_1(x)=1+x+x^2$ and initial state $\{1,  1\} $.
Next, consider the register $R_2$ with characteristic polynomial $p_2(x)=1+x+x^3$ and initial state $\{1, 1, 1\}$.
Then the shrunken-sequence can be computed as follows:
\[
\begin{array}{cccccccccccccccccccccc}
\{a_i\}:& 1&1&0&1&1&0&1&1&0&1&1&0&1&1&0&1&1&0&1&1&0\\
\{b_i\}:&1&1&\xcancel{1}&0&0&\xcancel{1}&0&1&\xcancel{1}&1&0&\xcancel{0}&1&0&\xcancel{1}&1& 1&\xcancel{0}&0&1&\xcancel{0}\\
 \{s_j\}:   & \pmb{1}&\pmb{1} & &\pmb{0}&\pmb{0} & &\pmb{0}&\pmb{1} &&\pmb{1}& \pmb{0}&&\pmb{1}&\pmb{0} &&\pmb{1}&\pmb{1} & &\pmb{0}&\pmb{1} &\\\
\end{array}
\]
The shrunken sequence $\{s_j\}$ (in bold) has period $14$ and it is not difficult to check that its characteristic polynomial is $p(x)^2=(1+x^2+x^3)^2$. Therefore, its linear complexity is $LC = 6$.  \hfill $\blacksquare$
\end{example}

%
%
%

\subsection{Cellular Automaton}\label{sec:CA}
\textbf{Cellular automata} (CAs) were first introduced by von Neumann as simple models to study biological processes such as self-reproduction \cite{Neumann1966bk}.
An elementary one-dimensional CA consists of an arrangement of cells  (with binary contents in this work) where the value of each cell evolves deterministically according to a set of rules involving its $k$ nearest neighbours.
Thus, the state of the cell in position $i$ at time $t+1$, denoted by $x_{i}^{t+1}$, depends on the state of the $k$ closest cells at time $t$.
If these rules are composed exclusively of XOR operations, then the CA is \textbf{linear}.
There are several types of CA: \textbf{null} (null cells are supposed to be adjacent to extreme cells) or \textbf{periodic} (extreme cells are adjacent), \textbf{regular} (every cell uses the same updating rule) or \textbf{hybrid} (different rules are applied to distinct cells).

The rules considered in this work are:
\bigskip

\begin{minipage}[]{0.4\textwidth}
\centering
\resizebox{7cm}{!} {
$
\begin{array}{l}
\text{\textbf{Rule 102:} }x_{i}^{t+1}=x_{i}^{t}+x_{i+1}^{t}\\
\\
\begin{array}{|c|c|c|c|c|c|c|c|}\hline
111 & 110 & 101 & 100 & 011 & 010 & 001 & 000\\ \hline
0  &  1  &  1  & 0   & 0   &  1  &  1  & 0 \\ \hline
\end{array}\\
\end{array}
$
}
\end{minipage}
\begin{minipage}[]{0.5\textwidth}
\centering
\resizebox{7cm}{!} {
$
\begin{array}{l}
\text{\textbf{Rule 60:} }x_{i}^{t+1}=x_{i-1}^{t}+x_{i}^{t}\\
\\
\begin{array}{|c|c|c|c|c|c|c|c|}\hline
111 & 110 & 101 & 100 & 011 & 010 & 001 & 000\\ \hline
0  &  0  &  1  & 1   & 1   &  1  &  0  & 0 \\ \hline
\end{array}\\
\end{array}
$
}
\end{minipage}
\bigskip

Recall that the numbers $01100110$ and $00111100$ are the binary representations of 102 and 60, respectively. This is the reason why they are called rule 102 and
rule 60.

Given a CA of length $l$ and initial state of the same length, it generates $l$ (vertical) sequences.
In Table~\ref{tab:1}, we have a regular, periodic, 102-CA, with initial state
$\{1, 0,  1,  0,  1,  1,  0,  0,  0,  1,  1,  1,  0,  1 \}$ that generates $14$ (vertical) sequences.

\begin{table}[!t]
\centering
\begin{scriptsize}
\caption{CA that generates the shrunken sequence in Example~\ref{ex:1}} \label{tab:1}
\begin{tabular}{|c|c|c|c|c|c|c|c|c|c|c|c|c|c|}
\hline
\bfseries 102 & \bfseries 102 & \bfseries 102 & \bfseries 102 & \bfseries 102 & \bfseries 102 & \bfseries 102 & \bfseries 102 & \bfseries 102 & \bfseries 102 & \bfseries 102 & \bfseries 102 & \bfseries 102 & \bfseries 102 \\
\hline\hline
\textbf{1} & 0 & \circled{1} & \circled{0} & \squared{1} & \squared{1} & \diamonded{0} & \diamonded{\textcolor{red}{0}} & 0 & 1 & \textcolor{green}{1} & 1 & 0 & 1  \\
\textbf{1} & 1 & 1 & 1 & 0 & 1 & 0 & 0 & 1 & 0 & 0 & 1 & 1 & 0\\
\diamonded{\textbf{0}} & \diamonded{\textcolor{red}{0}} & 0 & 1 & 1 & 1 & 0 & 1 & 1 & 0 & 1 & 0 & 1 & 1  \\\
\textbf{0} & 0 & 1 & 0 & 0 & 1 & 1 & 0 & 1 & 1 & 1 & 1 & 0 & 1  \\
\textbf{0} & 1 & 1 & 0 & 1 & 0 & 1 & 1 & 0 & 0 & 0 & 1 & 1 & 1  \\
\textbf{1} & 0 & 1 & 1 & 1 & 1 & 0 & 1 & 0 & 0 & 1 & 0 & 0 & 1  \\
\squared{\textbf{1}} & \squared{1} & 0 & 0 & 0 & 1 & 1 & 1 & 0 & 1 & 1 & 0 & 1 & 0  \\
\textbf{0} & 1 & 0 & 0 & 1 & 0 & 0 & 1 & 1 & 0 & 1 & 1 & 1 & 1  \\
\textbf{\textcolor{green}{1}} & 1 & 0 & 1 & 1 & 0 & 1 & 0 & 1 & 1 & 0 & 0 & 0 & 1  \\
\textbf{0} & 1 & 1 & 0 & 1 & 1 & 1 & 1 & 0 & 1 & 0 & 0 & 1 & 0  \\
\circled{\textbf{1}} & \circled{0} & 1 & 1 & 0 & 0 & 0 & 1 & 1 & 1 & 0 & 1 & 1 & 0  \\
\textbf{1} & 1 & 0 & 1 & 0 & 0 & 1 & 0 & 0 & 1 & 1 & 0 & 1 & 1  \\
\textbf{0} & 1 & 1 & 1 & 0 & 1 & 1 & 0 & 1 & 0 & 1 & 1 & 0 & 0 \\
\textbf{1} & 0 & 0 & 1 & 1 & 0 & 1 & 1 & 1 & 1 & 0 & 1 & 0 & 0  \\
\hline
\end{tabular}
\end{scriptsize}
\end{table}

\section{Modelling the shrunken sequence}\label{sec:mod}
In this section, we underline some properties of the shrunken sequence, which will be used in the algorithm  proposed in Section~\ref{sec:cryp}.
From now on, we consider two registers $R_1$ and $R_2$, with characteristic polynomials $p_1(x)$, $p_2(x)\in\mathbb{F}_2[x]$, lengths $L_1$ and $L_2$ ($\gcd(L_1,L_2)=1$)
and the periods of their corresponding m-sequences are $T_1=2^{L_1}-1$ and $T_2=2^{L_2}-1$, respectively.
Besides, the m-sequences generated by both registers are denoted by $\{a_i\}$ and $\{b_i\}$ ($i\geq 0$), respectively.
Since only the 1s of $\{a_i\}$ generate shrunken sequence bits, we assume without loss of generality that $a_0=1$.
As stated before, the shrunken sequence $\{s_j\}$ ($j\geq 0$) generated by both registers has period $T=2^{L_1-1}(2^ {L_2}-1)$ and characteristic polynomial $p(x)^L$, with $2^{L_1-2}< L \leq 2^ {L_1-1}$.

\subsection{Properties of the shrunken sequence}

If the shrunken sequence $\{s_j\}$ is decimated by  distance $d=2^{L_1-1}$ starting at position $s_i$, $i=0, 1, \ldots, d-1$, then we obtain $d$ m-sequences denoted by $\{s_{d\cdot j +i}\}$, for $i=0, 1, \ldots, d-1$ and $j\geq 0$.
\color{black}
Such sequences are called the \textbf{interleaved m-sequences} of the shrunken sequence and its characteristic polynomial  is again $p(x)$.
The polynomial $p(x)$ can be computed as follows,
\[
p(x)=(x+\alpha^{T_1})(x+\alpha^{2T_1})(x+\alpha^{4T_1})\cdots (x+\alpha^{2^{L_2-1}T_1}),
\]
where $\alpha$ is root of $p_2(x)$ (see \cite[Theorem. 3.3]{Cardell2016f}).

Let us see an illustrative example.

\begin{example}\label{ex:3}
Consider two registers with characteristic polynomials $p_1(x)=1+x+x^3$ and $p_2(x)=1+x+x^4$ and initial states $\{1, 0, 0\}$ and $\{1, 0, 0, 0\} $, respectively.
The shrunken sequence given by
\[\{
1   0   0   0\  1   1   1   1 \  1   0   1   0 \  0   0   0   1  \ 1   0   0   1\
   0   1   1   0  \ 1   1   0   0 \  1   1   0   1 \  0   1   0   0  \ 0   0   1   0 \
   1   1   1   0  \ 0   0   1   1  \ 0   1   1   1  \ 0   1   0   1  \ 1   0   1   1 \ldots\}
\]
has period $T=60$ and characteristic polynomial $p(x)^4=(1+x^3+x^4)^4$.
If we decimate the shrunken sequence by $4$, then we find that the shrunken sequence is composed of $4$ m-sequences:

\begin{equation}\label{eq:1}
\begin{array}{ccccccccccccccccc}
&&b_0 & b_7 & b_{14} & b_6 & b_{13} & b_{5} & b_{12} & b_4 & b_{11} & b_3 & b_{10} & b_2 & b_{9} & b_1 & b_8\\
&&\uparrow &\uparrow &\uparrow &\uparrow &\uparrow &\uparrow &\uparrow &\uparrow &\uparrow &\uparrow &\uparrow &\uparrow &\uparrow &\uparrow &\uparrow\\
\{s_{4j}\}&\rightarrow&\textbf{1} & \textbf{1} & \textbf{1} & \diamonded{\textbf{0}} & \textbf{1} & \squared{\textbf{0}} & \textbf{1} & \textbf{1} & \textbf{0} & \circled{\textbf{0}} & \textbf{1} & \textbf{0} & \textbf{0} & \textbf{0} & \textbf{1}\\
\{s_{4j+1}\}&\rightarrow&\circled{0} & 1 & 0 & 0 & 0 & 1 & 1 & 1 & 1 & 0 & 1 & 0 & 1 & 1 & 0\\
\{s_{4j+2}\}&\rightarrow&\squared{0} & 1 & 1 & 0 & 0 & 1 & 0 & 0 & 0 & 1 & 1 & 1 & 1 & 0 & 1\\
\{s_{4j+3}\}&\rightarrow&\diamonded{0} & 1 & 0 & 1 & 1 & 0 & 0 & 1 & 0 & 0 & 0 & 1 & 1 & 1 & 1\\
\end{array}
\end{equation}

These four m-sequences $\{s_{4j+i}\}$ $(0 \leq i < 4)$ have the same characteristic polynomial $p(x)=1+x^3+x^4$, thus all of them are shifted versions of one single m-sequence. This shift depends on
the positions of the $1$s in the m-sequence $\{a_i\}$ (see Section~\ref{sec:r2}). The bits $\circled{0}$, $\squared{0}$ and $\diamonded{0}$ in the sequence $\{s_{4j}\}$ represent the starting point of the sequences $\{s_{4j+i}\}$ $(1\leq i \leq 3)$, respectively. \hfill $\blacksquare$
\end{example}

\subsubsection{Relation between register $R_2$ and the shrunken sequence}
Given the shrunken sequence $\{s_j\}$, the following results help us to find the m-sequence $\{b_i\}$ generated by register $R_2$.

\begin{proposition}\label{prop:1}
Let $\delta \in \{1,2,3, \ldots, T_2-1\}$, such that $T_1\delta=1\text{ mod } T_2$.
If the first interleaved m-sequence is decimated by  distance $\delta$, then the resultant sequence is $\{b_i\}$.
 \end{proposition}
 \begin{proof}
According to the properties of the m-sequences,  $\{a_i\}$ has $2^{L_1-1}$ ones in the first $T_1$ bits \cite{Golomb1982bk}.
Besides, we know that the m-sequence $\{a_i\}$ decimates $\{b_i\}$ to obtain the shrunken sequence.
Therefore, the first interleaved m-sequence of the shrunken sequence is $\{b_0, b_{T_1}, b_{2T_1}, \ldots, b_{(T_2-1)T_1}\}$.
 If this sequence is decimated by  distance $\delta$, then the following sequence is obtained:
 $
 \{b_0, b_{\delta T_1}, b_{2\delta T_1}, \ldots, b_{(T_2-1)\delta T_1}\}.
 $
 We know that $T_1\delta =1\text{ mod } T_2$ and, therefore, the sequence can be seen as
  $
 \{b_0, b_{1}, b_{2 }, \ldots, b_{(T_2-1)}\},
 $
 which is the m-sequence generated by the register $R_2$.$\hfill\square$
 \end{proof}

The previous proposition leads us to the following results.

 \begin{corollary}
If the shrunken sequence is decimated by  distance $2^{L_1-1}\delta$, then the m-sequence $\{b_i\}$ is obtained.
\end{corollary}

\begin{corollary}\label{cor:1}
If the primitive polynomials $p_1(x)$, $p_2(x)\in \mathbb{F}_2[x]$ have degrees $L_1$ and $L_1+1$, respectively, then $\delta=T_2-2$.
\end{corollary}

\begin{proof}
We proceed with the following computations:
\begin{align*}
(T_2-2)T_1=(2^{L_1+1}-3)(2^{L_1}-1)=2^{2L_1+1}-3\cdot 2^{L_1}-2^{L_1+1}+3 =2^{2L_1+1}-(2^{L_1+1}-1)\\
-(2^{L_1+1}-1)-2^{L_1}+1=2^{L_1}(2^{L_1+1}-1)-2(2^{L_1+1}-1)+1.
\end{align*}
Since $T_2=2^{L_1+1}-1$, then $(T_2-2)T_1= 1 \text{ mod }T_2$.
Consequently, we have that $\delta=T_2-2$.\hfill$\square$
\end{proof}

In Example~\ref{ex:3}, we had that $L_1=3$ and $L_2=4$.
In this case, we can apply Corollary~\ref{cor:1} and $\delta=T_2-2=13$.
If we decimate the first interleaved m-sequence by $13$ (see expression~(\ref{eq:1})), then we obtain $\{b_i\}$, the m-sequence generated by $p_2(x)=1+x+x^ 4$.
In this case $\{b_i\}=\{1 0 0 0 1  0 0 1 1 0 1 0 1 1 1\}$.

\subsubsection{Relation between register $R_1$ and the shrunken sequence}\label{sec:r2}

In this section we analyse  how to recover the m-sequence $\{a_i\}$ from the shrunken sequence $\{s_j\}$.
Assume the first interleaved m-sequence is denoted by $\{v_i\}$.
Since the other interleaved m-sequences are the same but shifted, we assume they have the form $\{v_{d_1+i}\}, \{v_{d_2+i}\}, \ldots, \{v_{d_{2^{L_1-1}-1}+i}\}$ for some $d_i\in\{0, 1, 2, \ldots, 2^{L_2-2}\}$:

$$\begin{array}{rccccc}
\{v_i\}: &	v_{0}     & v_{1}       & v_{2}       & \ldots &  v_{T_2-1} \\
 \{v_{d_1+i}\} :&	v_{d_1}     & v_{d_1+1}     & v_{d_1+2}     & \ldots & v_{d_1+T_2-1} \\
 \{v_{d_2+i}\}:	&v_{d_2}     & v_{d_2+1}     & v_{d_2+2}     & \ldots &  v_{d_2+T_2-1}\\
 	\multicolumn{1}{c}{\phantom{aaaa}\vdots}&\vdots    & \vdots        &  \vdots       &        & \vdots \\
\{v_{d_{2^{L_1-1}-1}+i}\}:& 	v_{d_{2^{L_1-1}-1}} & v_{d_{2^{L_1-1}-1}+1} & v_{d_{2^{L_1-1}-1}+2} & \ldots &  v_{d_{2^{L_1-1}-1}+T_2-1}\\
\end{array}$$

In order to illustrate this idea, consider again Example~\ref{ex:3}.
In this case, we had four interleaved m-sequences
\[
\{s_{4j}\}=\{v_i\}, \quad \{s_{4j+1}\}=\{v_{d_1+i}\}, \quad \{s_{4j+2}\}=\{v_{d_2+i}\} \quad \text{and} \quad \{s_{4j+3}\}=\{v_{d_3+i}\}.
\]

In is well known that a maximum-length LFSR of length $L$, produces an m-sequence with $2^{L-1}$ ones in the first period.
Now, we are ready to introduce the following result.

\begin{proposition}
Let $\{0,i_1, i_2, \ldots, i_{2^{L_1-1}-1}\}$ be the set of positions of the $1$s in the m-sequence $\{a_i\}$ in its first period.
Therefore, $d_k=\delta \cdot i_k \text{ mod } (2^{L_1-1}-1)$, for $k=1, 2,\ldots, 2^{L_1-1}-1$, where $\delta$ has the form given in Proposition~\ref{prop:1}.
\end{proposition}

In Example~\ref{ex:3}, we had four interleaved m-sequences $\{v_i\}$, $\{v_{i+d_1}\}$, $\{v_{i+d_2}\}$ and $\{v_{i+d_3}\}$.
It is easy to check, from  expression~(\ref{eq:1}), that  $d_1=9$, $d_2=5$ and $d_3=3$.
In this case, $T_1=7$ and $T_2=15$, then, according to Corollary~\ref{cor:1}, $\delta=13$.
%
With this information, we can compute the positions of the $1$s in $\{a_i\}$ ($i_0=0$, without loss of generality):
\begin{align*}
13\cdot i_1 = 9 \text{ mod } 15\rightarrow i_1=3\\
13\cdot i_2 = 5 \text{ mod } 15\rightarrow i_2=5\\
13\cdot i_3 = 3 \text{ mod } 15\rightarrow i_3=6
\end{align*}
Therefore, the set of positions is given by $\{0, 3,5,6 \}$ and then the m-sequence is $\{a_i\}=\{1, 0, 0, 1 ,0, 1, 1\}$.\hfill $\blacksquare$

\color{black}
\subsection{Characterization of the CA}\label{sec:CH:CA}

In \cite[Theorem 3.10]{Cardell2016f}, the authors showed that there exists a 102-CA of length $L=\frac{T}{\gcd(2^{L_2}-1,D)}$ that generates the shrunken sequence as one of the (vertical) sequences.
We remind that $T$ is the period of the shrunken sequence and   $D=\ensuremath{\mathcal{Z}}_{\alpha}(1)$, where $\ensuremath{\mathcal{Z}}_{\alpha}(1)$ is the Zech logarithm of $1$ on basis $\alpha$ \cite{Huber1990} (see also Definition~\ref{def:1} in Section~\ref{sec:zech}).

Furthermore, it was shown that there are $2^{L_1-1}$ different sequences that appear repeated several times along the CA.
Such sequences have the same characteristic polynomial $p(x)^L$ as that of the shrunken sequence \cite[Theorem 3.9]{Cardell2016f}.
We call these sequences the \textbf{companion sequences} of the shrunken sequence.

\begin{example}\label{ex:2}
In Example~\ref{ex:1}, we obtained a shrunken sequence of period $T=14$ and characteristic polynomial $p(x)^2=(1+x^2+x^3)^2$.
In Table~\ref{tab:1}, it is possible to see how to obtain this sequence as one of the output (vertical) sequences of a 102-CA of length 14.
It is possible to check that there  is another sequence apart from the shrunken sequence that appears shifted along the CA:
$\{0\ 1\ 0\ 0\ 1\ 0\ 1\ 1\ 1\ 1\ 0\ 1\ 1\ 1\ 0\}$.
Both sequences appear seven times, but shifted each time $D=2\cdot \ensuremath{\mathcal{Z}}_{\alpha}(1)=10$ positions.
\hfill $\blacksquare$
\end{example}


\subsection{Companion sequences}\label{sec:com}
We have seen that if we locate the shrunken sequence in the zero column of the CA, $2^{L_1-1}$ different sequences are generated, including the shrunken sequence.
All these sequences have the same characteristic polynomial $p(x)^L$ and are composed by interleaving $2^{L_1-1}$ m-sequences with characteristic polynomial~$p(x)$~\cite{Cardell2016f}.

In Example~\ref{ex:2}, two sequences were generated by the CA, the shrunken sequence and another sequence with the same period and the same characteristic polynomial $p(x)^2=(1+x^2+x^3)^2$.
If we decimate both sequences by $2$, we can see that both sequences are composed of two m-sequences whose characteristic polynomial is $p(x)=1+x^2+x^3$.
This means, that all the m-sequences are the same but shifted (see Table~\ref{tab:70})

Consequently, we can deduce that the fact of knowing some bits of the companion sequences of the CA can help us to recover parts of the shrunken sequence.
More precisely, we can recover the first interleaved m-sequence of the shrunken sequence using the  interleaved m-sequences of the same shrunken sequence and the interleaved m-sequences of the companion sequences.

Consider a linear $102$-CA  and  consider that the sequence in the zero column is the shrunken sequence $\{s_i\}$.
The form of the corresponding companion sequences is computed in Table~\ref{tab:3}.
It is not difficult to check that the companion sequences in columns whose indices are   $2^j$, for $j= 0, 1, 2, \ldots$, have the form $\{s_i+s_{i+2^j}\}$.
In fact, the general form of these columns can be found in  \cite{Cardell2016c}.

\begin{table}
\scriptsize
\centering
\caption{General $102$-CA \label{tab:3}}
\[
\begin{array}{|c|c|c|c|c|c|c|c|}\hline
\textbf{102} & \textbf{102} & \textbf{102} & \textbf{102} & \textbf{102} & \ldots  & \textbf{102} & \ldots \\\hline\hline
s_0 & s_0+s_1 & s_0+s_2 & s_0+s_1+s_2+s_3 & s_0+s_4 & \ldots & s_0+s_8  & \ldots\\
s_1 & s_1+s_2 & s_1+s_3 & s_1+s_2+s_3+s_4 & s_1+s_5 & \ldots & s_1+s_9 & \ldots\\
s_2 & s_2+s_3 & s_2+s_4 & s_2+s_3+s_4+s_5 & s_2+s_6 & \ldots & s_2+s_{10} & \ldots\\
s_3 & s_3+s_4 & s_3+s_5 & s_3+s_4+s_5+s_6 & s_3+s_7 & \ldots & s_3+s_{11} & \ldots\\
\vdots & \vdots & \vdots & \vdots & \vdots &  & \vdots &   \\\hline
\end{array}
\]
\end{table}

Let us denote the interleaved m-sequences of the shrunken sequence by $\{v_{d^0_0+i}\}$, $\{v_{d^0_1+i}\}$, $\{v_{d^0_2+i}\}$, $\ldots, \{v_{d^0_{2^{L_1-1}-1}+i}\}$, $i\geq 0$, where $d_0^0=0$.
Remember that the positions $d^0_k$ depend on the location of the 1s in the m-sequence $\{a_i\}$ generated by the first register $R_1$ (see Section~\ref{sec:r2}).

Let us denote the interleaved m-sequences of the first companion sequence in the CA  by $\{v_{d_0^1+i}\}$, $\{v_{d_1^1+i}\}$, $\{v_{d_2^1+i}\}$, $\ldots, \{v_{d_{2^{L_1-1}-1}^1+i}\}$, $i\geq 0$. 
We can compute  these new positions using the definitions of   rule 102 and Zech logarithm as follows
	\begin{align*}
		 &d_k^1=\zech{\alpha}{d^0_{k}-d^0_{k+1}}+d^0_{k+1},    k=0,1, \ldots, 2^{L_1-1}-2,\\
		 &d_{2^{L_1-1}-1}^ 1=\zech{\alpha}{d^0_{2^{L_1-1}-1}-1}+1.
	 \end{align*}

Similarly, we can compute the shifts positions for the $j$-th companion sequence, $j=1,2, \ldots, L-1$ as:
	\begin{align}
		\label{eq:3a} &d_k^j=\zech{\alpha}{d_{k}^{j-1}-d_{k+1}^{j-1}}+d_{k+1}^{j-1},    k=0,1, \ldots, 2^{L_1-1}-2,\\
\nonumber 	 &d_{2^{L_1-1}-1}^j=\zech{\alpha}{d^{j-1}_{2^{L_1-1}-1}-1}+1.
	 \end{align}
	
Notice that the companion sequences in columns $t\cdot 2^{L_1-1}$, with $t=1, 2, \ldots, L/(2^{L_1-1})-1$ in the CA, are again the shrunken sequence, but starting in positions $t\cdot D\cdot 2^{L_1-1}$, respectively \cite[Theorem 3.8]{Cardell2016f}.
Moreover, the companion sequence  in the column  $t\cdot 2^{L_1-1}+m$
for $m=1, 2, \ldots, 2^{L_1-1}-1$ and $t=0, 1, \ldots, L/(2^{L_1-1})-1$
is the same as the companion sequence in the $m$-th column starting in position $t\cdot D\cdot 2^{L_1-1}$.
Therefore, we have that:
\[
d_k^{t\cdot 2^{L_1-1}+m}=d_k^{m}+t\cdot  D \mod (2^{L_2}-1)
\]
for $k=0, 1, \ldots, 2^{L_1-1}-1$, $m=0, 1, \ldots, 2^{L_1-1}-1$ and $t=0, 1, \ldots, L/(2^{L_1-1})-1$.


This means that, the positions $d_k^j$ for the  companion sequence  in the $j$-th column  with $j\geq 2^{L_1-1}$ can be computed easily using the positions $d_i^s$, with $0\leq s < 2^{L_1-1}$ (without logarithms).

%
%
%

\begin{example}
Consider  the shrunken sequence in Example~\ref{ex:1}.
The CA in Table~\ref{tab:1} generates two sequences, the shrunken sequence and one companion sequence.
Both sequences are composed of interleaved m-sequences (see Table~\ref{tab:70}).

Let us consider the sequence in the 7-th column of the CA.
In this particular case, we have that $t=3$, $D=5$ and $2^{L_1-1}=2$.
Therefore, this sequence is the same sequence  as the companion sequence but starting in position $t\cdot D\cdot 2^{2^{L_1-1}} \bmod 14= 2$ (see red bits in Table~\ref{tab:1}).
Furthermore, the two interleaved m-sequences of this sequence are the same as the first interleaved m-sequence of the shrunken sequence (Table~\ref{tab:70a})  starting in positions:
\begin{align*}
d_0^7=d_0^1+D\cdot 3 \bmod 7= 2\quad \text{and}\quad
d_1^7=d_1^1+D\cdot 3 \bmod 7 =1,   \text{respectively}.
\end{align*}

Let us consider the column in the 10-th position, that is, $t=5$. Then, this sequence is the shrunken sequence, but starting in position $t\cdot D\cdot 2^{L_1-1}\bmod 14 =8$ (see the green bits in Table~\ref{tab:1}).
Furthermore, the two interleaved m-sequences this sequence is composed of are the same as the first interleaved m-sequence of the shrunken sequence (Table~\ref{tab:70a}) starting in positions:
\begin{align*}
d_0^{10}=d_0^0+D\cdot 5 \bmod 7= 4 \quad \text{and} \quad 
d_1^{10}=d_1^0+D\cdot 5 \bmod 7 =2, \text{respectively.}
\end{align*}
\hfill $\blacksquare$
\end{example}





\begin{table}
\caption{Interleaved m-sequences of the shrunken sequence and the companion sequences of Example~\ref{ex:3} \label{tab:70}}
\centering\scriptsize
\subfloat[][\label{tab:70a}]{
\begin{tabular}{cc}
\underline{1}&1\\
0&0\\
0&\circled{1}\\
1&0\\
1&0\\
1&1\\
0&1\\
\end{tabular} }
\qquad
\subfloat[][\label{tab:70b}]{
\begin{tabular}{cc}
0&\diamonded{1}\\
0&0\\
1&0\\
1&1\\
1&1\\
0&1\\
\squared{1}&0\\
\end{tabular}
}
\qquad
\subfloat[][\label{tab:70c}]{
\begin{tabular}{cc}
0& 0\\
1& 0\\
1& 1 \\
1& 1\\
0& 1 \\
1& 0 \\
0& 1\\
\end{tabular} }
\qquad
\subfloat[][\label{tab:70d}]{
\begin{tabular}{cc}
1 &0\\
1 &1\\
0 &1\\
1 &1\\
0 &0\\
0 &1\\
1 &0\\
\end{tabular}
 }
\end{table}

\section{Zech's logarithm}\label{sec:zech}
Zech logarithms are named after Julius Zech which published a table of these type logarithms (which he called \textit{addition logarithms}) for doing arithmetic in $\Zset_p$.
These logarithms are also called as Jacobi logarithms  after C. G. J. Jacobi who used them for number theoretic investigations \cite{Jacobi1846}.

Assume we are working over the finite field $\Fset_{q}$, where $q=p^m$, with $p$ prime and $m$ a positive integer.
We now introduce the definition of Zech logarithm.

 \begin{definition}\label{def:1}
Let $\alpha\in\mathbb{F}_{q}$ be a primitive element.
The \textbf{Zech logarithm} with basis $\alpha$  is the application $\ensuremath{\mathcal{Z}}_{\alpha}:\mathbb{Z}_{q} \rightarrow \mathbb{Z}_{q}^*\cup \{ \infty\} $, such that each element $t\in\mathbb{Z}_{q}$ corresponds to $\ensuremath{\mathcal{Z}}_{\alpha}(t)$, attaining $1+\alpha^t=\alpha^{\ensuremath{\mathcal{Z}}_{\alpha}(t)}$.
\end{definition}

\begin{example}
Let $\alpha \in \Fset_{2^3}$ a  root of the primitive polynomial $p(x)=1+x+x^3$.
Then:
\begin{align*}
\alpha^3&=1+\alpha \rightarrow \ensuremath{\mathcal{Z}}_{\alpha}(1)=3\\
\alpha^4&= \alpha+\alpha^2\\
\alpha^5&=\alpha^2+\alpha^3=1+\alpha+\alpha^2=1+\alpha^4 \rightarrow \ensuremath{\mathcal{Z}}_{\alpha}(4)=5\\
\alpha^6&=\alpha+\alpha^2+\alpha^3=1+\alpha^2\rightarrow \ensuremath{\mathcal{Z}}_{\alpha}(2)=6\\
\end{align*}
Next we find the complete Zech logarithm table for $\Fset_{2^3}$:
\[
\begin{array}{|c|c|c|c|c|c|c|}\hline
x&1&2&3&4&5&6\\ \hline
\ensuremath{\mathcal{Z}}_{\alpha}(x)&3&6&1&5&4&2\\\hline
\end{array} \quad \hfill \blacksquare
\]
\end{example}

Zech logarithms are discrete logarithms and they are, thus, hard to compute.
Now, we are going to show some of their properties that can be used to easily compute them with trivial operations.
These results can be found in  \cite{Huber1990}.

\begin{proposition}\label{prop:2}
Given the Zech logarithm defined as in Definition~\ref{def:1}.
The following properties follow easily:
\begin{enumerate}
\item[a)] $\ensuremath{\mathcal{Z}}_{\alpha}(q-1-x)=\ensuremath{\mathcal{Z}}_{\alpha}(x)-x \bmod {(q-1)}$
\item[b)] $\ensuremath{\mathcal{Z}}_{\alpha}(p\,x)=p\,\;\ensuremath{\mathcal{Z}}_{\alpha}(x)\bmod{(q-1)}$\label{prop:2b}
\item[c)] $\ensuremath{\mathcal{Z}}_{\alpha}(0)=-\infty$, for $p=2$
\item[d)] $\ensuremath{\mathcal{Z}}_{\alpha}(\frac{q-1}{2})=-\infty$, for $p\not =2$
\end{enumerate}

\end{proposition}

\begin{remark}\label{rem:1}
Notice that for fields of characteristic two we have that
$\ensuremath{\mathcal{Z}}_{\alpha}(n) = m$ implies that $\ensuremath{\mathcal{Z}}_{\alpha}(m) = n $.
\end{remark}

Consider now the notion of cyclotomic coset $\bmod{(q-1)}$ given in \cite{Golomb1982bk}.

\begin{definition}
Let $\Fset_{2^L}$ denote the Galois field of $2^L$ elements.
An equivalence relation $R$ is defined on its elements $\alpha, \beta \in \Fset_q$ such as follows:
$\alpha \, R \, \beta $ if there exists an integer $j$, $0 \leq j \leq L-1$, such that
\[
 2^j \cdot \alpha = \beta  \bmod  (2^L -1) .
\]
The resultant equivalence classes into which $\Fset_{2^L}^*$ is partitioned are called the \emph{cyclotomic cosets} modulo $2^L-1$.
\end{definition}
The leader element  of every coset is the smallest integer in such an equivalence class.

According to \cite{Huber1990}, Zech logarithms  map cosets onto cosets of the same size, that is, $\ensuremath{\mathcal{Z}}_{\alpha}:C_{s_1}\rightarrow C_{s_2}$, where $\card{C_{s_1}}=\card{C_{s_2}}$.

On the other hand, we consider  the mapping $I:N_q\rightarrow N_q$, with $I(x)=q-1-x$, where $N_q=\{0, 1, 2, \ldots, q-2\}\cup \{-\infty\}$.
According to \cite{Huber1990}, $\ensuremath{\mathcal{Z}}_{\alpha}(I(x))=\ensuremath{\mathcal{Z}}_{\alpha}(x)-x\bmod{(q-1)}$
and
\begin{equation}\label{eq:2}
\ensuremath{\mathcal{Z}}_{\alpha}^{-1}(\ensuremath{\mathcal{Z}}_{\alpha}(x)-x)=I(x).
\end{equation}

Like $\ensuremath{\mathcal{Z}}_{\alpha}(x)$, the application $I(x)$ also maps a cyclotomic coset onto another of the same size.
If $\ensuremath{\mathcal{Z}}_{\alpha}(x)$ is known, then the value of $\ensuremath{\mathcal{Z}}_{\alpha}(I(x))$ can be computed.
For fields with characteristic two, $\ensuremath{\mathcal{Z}}_{\alpha}^{-1}(x)=\ensuremath{\mathcal{Z}}_{\alpha}(x)$ and the computations are particularly simple.

\begin{example}
Consider the field $\Fset_{2^5}$, constructed with the primitive polynomial $p(x)=1+x^2+x^5$.
There are six cyclotomic cosets given by:
\begin{align*}
C_1&=\{1,2,4,8,16\}\quad &C_7=\{7,1,14,25,19\}\\
C_3&=\{3,6,12,24,17\}\quad & C_{11}=\{11,22,13,26\}\\
C_5&=\{5,10,20,9,18\}\quad & C_{15}=\{15,30,29,27\}\\
\end{align*}
From $p(x)$ we know that $\ensuremath{\mathcal{Z}}_{\alpha}(2)=5$.
Since we are working over a field with characteristic two and according to b) in Proposition~\ref{prop:2}, we can compute the logarithms of the other elements of $C_2$ and $C_5$:
\begin{align*}
\ensuremath{\mathcal{Z}}_{\alpha}(2)=5 &\rightarrow \ensuremath{\mathcal{Z}}_{\alpha}(5)=2\\
\ensuremath{\mathcal{Z}}_{\alpha}(4)=2\ensuremath{\mathcal{Z}}_{\alpha}(2)=10  &\rightarrow \ensuremath{\mathcal{Z}}_{\alpha}(10)=4\\
\ensuremath{\mathcal{Z}}_{\alpha}(8)=2\ensuremath{\mathcal{Z}}_{\alpha}(4)=20  &\rightarrow \ensuremath{\mathcal{Z}}_{\alpha}(20)=8\\
\ensuremath{\mathcal{Z}}_{\alpha}(16)=2\ensuremath{\mathcal{Z}}_{\alpha}(8)=9  &\rightarrow \ensuremath{\mathcal{Z}}_{\alpha}(9)=16\\
\ensuremath{\mathcal{Z}}_{\alpha}(1)=2\ensuremath{\mathcal{Z}}_{\alpha}(16)=18 & \rightarrow \ensuremath{\mathcal{Z}}_{\alpha}(18)=1\\
\end{align*}

Let us now use the map $I$ to obtain the logarithms for elements in $C_3$ and $C_{15}$.
According to the definition of this map, it is easy to compute $I(2)=29$.
Then, according to (\ref{eq:2}), we know that $\ensuremath{\mathcal{Z}}_{\alpha}(29)=3$.
Using the same method as above, we can compute every logarithm in $C_3$ and $C_{15}$.

Equally, we have $I(3)=28$ and according to equation (\ref{eq:2}), $\ensuremath{\mathcal{Z}}_{\alpha}(28)=26$.
Therefore, we can compute every logarithm in cosets $C_{11}$ and $C_7$.

Using these properties, we can compute every logarithm in $\Fset_{2^5}$, with very simple operations $\bmod{\,31}$.
The complete Zech logarithm table for $\Fset_{2^5}$ can be found in Table~\ref{tab:4}. 
\hfill $\blacksquare$
\end{example}

\begin{table}
\caption{Zech logarithms for $\Fset_{2^ 5}$\label{tab:4}}
 \[
\begin{array}{|c|c||c|c||c|c||c|c|}\hline
x & \ensuremath{\mathcal{Z}}_{\alpha}(x)&x&\ensuremath{\mathcal{Z}}_{\alpha}(x)&x&\ensuremath{\mathcal{Z}}_{\alpha}(x)&x&\ensuremath{\mathcal{Z}}_{\alpha}(x)\\\hline
0&-\infty&8 &20 &16 &9 &24&15\\
1&18&     9 &16 &17 &30&25&21\\
2&5&      10& 4 &18 &1 &26&28\\
3&29&     11&19 &19 &11&27&6\\
4&10&     12&23 &20 &8 &28&26\\
5&2&      13&14 &21 &25&29&3\\
6&27&     14&13 &22 &7 &30&17\\
7&22&     15&24 &23 &12& &\\\hline

\end{array}
\]
\end{table}

Next, we find some minor results about Zech logarithms. 

\begin{proposition}\label{prop:3}
If $\beta_1=\zech{\alpha}{\alpha_1-\alpha_2}+\alpha_2$ and $\beta_2=\zech{\alpha}{\alpha_2-\alpha_3}+\alpha_3$, then:
\begin{enumerate}
\item $\beta_1=\zech{\alpha}{\alpha_2-\alpha_1}+\alpha_1$
\item $\zech{\alpha}{\alpha_1-\alpha_3}+\alpha_3=\zech{\alpha}{\beta_1-\beta_2}+\beta_2$ (for fields of characteristic equal to two)
\end{enumerate}
\end{proposition}
\begin{proof}
\begin{enumerate}
\item According to the definition of Zech logarithm:

$\alpha^{\beta_1}=\alpha^{\zech{\alpha}{\alpha_1-\alpha_2}+\alpha_2} =  (1+\alpha^{\alpha_1-\alpha_2})\alpha^{\alpha_2}=\alpha^{\alpha_1}+\alpha^{\alpha_2}=(1+\alpha^{\alpha_2-\alpha_1})\alpha^{\alpha_1}=\alpha^{\zech{\alpha}{\alpha_2-\alpha_1}+\alpha_1}$
\item Again, according to the definition of Zech logarithm:
\begin{align}
\label{a} \beta_1=\zech{\alpha}{\alpha_1-\alpha_2}+\alpha_2 \rightarrow  \alpha ^{\beta_1}=\alpha^{\alpha_1}+\alpha^{\alpha_2}\\
\label{b} \beta_2=\zech{\alpha}{\alpha_2-\alpha_3}+\alpha_3 \rightarrow \alpha ^{\beta_2}=\alpha^{\alpha_2}+\alpha^{\alpha_3}
\end{align}
Summing both equations (\ref{a}) and (\ref{b}), we get
$$  \alpha^{\beta_1}+\alpha^{\beta_2}=\alpha^{\alpha_1}+\alpha^{\alpha_3} \rightarrow \zech{\alpha}{\beta_1-\beta_2}+\beta_2=\zech{\alpha}{\alpha_1-\alpha_3}+\alpha_3$$
\end{enumerate}
\hfill $\square$
\end{proof}

\begin{remark}\label{rem:2}
\begin{enumerate}
\item The first part of Proposition~\ref{prop:3} implies that
\[
d_i^j=\zech{\alpha}{d_{i}^{j-1}-d_{i+1}^{j-1}}+d_{i+1}^{j-1}=\zech{\alpha}{d_{i+1}^{j-1}-d_{i}^{j-1}}+d_{i}^{j-1},
\]
and, thus, we can change the order of the computations if needed.
\item The second part of Proposition~\ref{prop:3} implies that
$$d_i^j=\zech{\alpha}{d_{i}^{j-1}-d_{i+1}^{j-1}}+d_{i+1}^{j-1}=\zech{\alpha}{d_{i}^{j-2}-d_{i+2}^{j-2}}+d_{i+2}^{j-2},$$
and, thus, we can choose the best positions for our computations.
\end{enumerate}
\end{remark}

The algorithm we will propose in Section~\ref{sec:cryp} makes use of the Zech logarithm to compute the positions given in~(\ref{eq:3a}).
Using the properties we have  seen in this section, we can reduce the complexity of the algorithm reducing the number of calculations.

\section{Recovering the shrunken sequence}\label{sec:cryp}
In this section, we use the CAs and their properties to recover the complete shrunken sequence.
We know that the shrunken sequence and the companion sequences appear several times along the same CA, with shift $D=2^{L_1-1}\ensuremath{\mathcal{Z}}_{\alpha}(1)$ (see Section~\ref{sec:CH:CA}).

In Example~\ref{ex:1}, we had a shrunken sequence of period 14.
In Table~\ref{tab:1}, we saw that there exists a CA of length 14 that produces this sequence in its leftmost column.
If we intercept the first 6 bits of the shrunken sequence, we can recover 21 elements in the CA (see the purple triangle in Table~\ref{tab:5}).
According to the properties of this CA, these sequences are repeated along the CA and, thus, we can recover the same number of bits in other positions (see the other triangles in Table~\ref{tab:5}).
In this case, the triangles of recovered bits overlap.
Therefore, we can recover completely all the elements in the CA and, thus, recover the complete shrunken sequence.

\begin{table}[!t]
\centering
\begin{scriptsize}
\caption{CA that generates the shrunken sequence in Example~\ref{ex:1}} \label{tab:5}
$$\begin{array}{|c|c|c|c|c|c|c|c|c|c|c|c|c|c|}\hline
102 & 102 &  102 & 102 & 102 & 102 & 102 &  102 & 102 & 102 & 102 & 102 &  102 &  102 \\ \hline\hline
\cellcolor[cmyk]{0.5,0.5,0,0}1 & \cellcolor[cmyk]{0.5,0.5,0,0}1 & \cellcolor[cmyk]{0.5,0.5,0,0}0 & \cellcolor[cmyk]{0.5,0.5,0,0}1 & \cellcolor[cmyk]{0.5,0.5,0,0}0 & \cellcolor[cmyk]{0.5,0.5,0,0}0 & \cellcolor[cmyk]{0,0.5,0.5,0}1 & \cellcolor[cmyk]{0,0.5,0.5,0}0 &\cellcolor[cmyk]{0,0.5,0.5,0} 0 &\cellcolor[cmyk]{0,0.5,0.5,0} 1 & \phantom{1} & \phantom{0} & \cellcolor[cmyk]{0,0.5,0,0}1 & \cellcolor[cmyk]{0,0.5,0,0}1  \\ \cline{6-6} \cline{10-10} \cline{14-14}
\cellcolor[cmyk]{0.5,0.5,0,0}0 &\cellcolor[cmyk]{0.5,0.5,0,0} 1 &\cellcolor[cmyk]{0.5,0.5,0,0} 1 &\cellcolor[cmyk]{0.5,0.5,0,0} 1 &\cellcolor[cmyk]{0.5,0.5,0,0} 0 & \phantom{1} &\cellcolor[cmyk]{0,0.5,0.5,0} 1 &\cellcolor[cmyk]{0,0.5,0.5,0} 0 &\cellcolor[cmyk]{0,0.5,0.5,0} 1 & \phantom{0} & \phantom{1} & \phantom{1} & \cellcolor[cmyk]{0,0.5,0,0}0 & \phantom{0} \\  \cline{5-5} \cline{9-14}
\cellcolor[cmyk]{0.5,0.5,0,0}1 & \cellcolor[cmyk]{0.5,0.5,0,0}0 &\cellcolor[cmyk]{0.5,0.5,0,0} 0 &\cellcolor[cmyk]{0.5,0.5,0,0} 1 & \phantom{1} & \phantom{0} & \cellcolor[cmyk]{0,0.5,0.5,0}1 & \cellcolor[cmyk]{0,0.5,0.5,0}1 & \cellcolor[cmyk]{0.5,0,0.5,0}1 & \cellcolor[cmyk]{0.5,0,0.5,0}1 & \cellcolor[cmyk]{0.5,0,0.5,0}0 & \cellcolor[cmyk]{0.5,0,0.5,0}1 & \cellcolor[cmyk]{0.5,0,0.5,0}0 & \cellcolor[cmyk]{0.5,0,0.5,0}0  \\  \cline{4-4} \cline{8-8} \cline{14-14}
\cellcolor[cmyk]{0.5,0.5,0,0}1 &\cellcolor[cmyk]{0.5,0.5,0,0} 0 &\cellcolor[cmyk]{0.5,0.5,0,0} 1 & \phantom{0} & \phantom{1} & \phantom{1} & \cellcolor[cmyk]{0,0.5,0.5,0}0 & \phantom{0} & \cellcolor[cmyk]{0.5,0,0.5,0}0 & \cellcolor[cmyk]{0.5,0,0.5,0}1 & \cellcolor[cmyk]{0.5,0,0.5,0}1 & \cellcolor[cmyk]{0.5,0,0.5,0}1 & \cellcolor[cmyk]{0.5,0,0.5,0}0 & \phantom{1}  \\ \cline{3-8}   \cline{13-13}
\cellcolor[cmyk]{0.5,0.5,0,0}1 &\cellcolor[cmyk]{0.5,0.5,0,0} 1 & \cellcolor[cmyk]{0,0,0.5,0}1 & \cellcolor[cmyk]{0,0,0.5,0}1 & \cellcolor[cmyk]{0,0,0.5,0}0 & \cellcolor[cmyk]{0,0,0.5,0}1 & \cellcolor[cmyk]{0,0,0.5,0}0 & \cellcolor[cmyk]{0,0,0.5,0}0 & \cellcolor[cmyk]{0.5,0,0.5,0}1 &\cellcolor[cmyk]{0.5,0,0.5,0} 0 & \cellcolor[cmyk]{0.5,0,0.5,0}0 & \cellcolor[cmyk]{0.5,0,0.5,0}1 & \phantom{1} & \phantom{0}\\ \cline{2-2} \cline{8-8} \cline{12-12}
\cellcolor[cmyk]{0.5,0.5,0,0}0 & \phantom{0} & \cellcolor[cmyk]{0,0,0.5,0}0 & \cellcolor[cmyk]{0,0,0.5,0}1 & \cellcolor[cmyk]{0,0,0.5,0}1 & \cellcolor[cmyk]{0,0,0.5,0}1 & \cellcolor[cmyk]{0,0,0.5,0}0 & \phantom{1} & \cellcolor[cmyk]{0.5,0,0.5,0}1 & \cellcolor[cmyk]{0.5,0,0.5,0}0 & \cellcolor[cmyk]{0.5,0,0.5,0}1 & \phantom{0} & \phantom{1} & \phantom{1}  \\ \cline{1-2} \cline{7-7} \cline{11-14}
\cellcolor[cmyk]{0.5,0,0,0}0 & \cellcolor[cmyk]{0.5,0,0,0}0 & \cellcolor[cmyk]{0,0,0.5,0}1 & \cellcolor[cmyk]{0,0,0.5,0}0 & \cellcolor[cmyk]{0,0,0.5,0}0 & \cellcolor[cmyk]{0,0,0.5,0}1 & \phantom{1} & \phantom{0} & \cellcolor[cmyk]{0.5,0,0.5,0}1 & \cellcolor[cmyk]{0.5,0,0.5,0}1 & \cellcolor[cmyk]{0.5,0,0,0}1 & \cellcolor[cmyk]{0.5,0,0,0}1 & \cellcolor[cmyk]{0.5,0,0,0}0 &\cellcolor[cmyk]{0.5,0,0,0} 1  \\ \cline{2-2} \cline{6-6} \cline{10-10}
\cellcolor[cmyk]{0.5,0,0,0}0 & \phantom{1} & \cellcolor[cmyk]{0,0,0.5,0}1 & \cellcolor[cmyk]{0,0,0.5,0}0 & \cellcolor[cmyk]{0,0,0.5,0}1 & \phantom{0} & \phantom{1} & \phantom{1} & \cellcolor[cmyk]{0.5,0,0.5,0}0 & \phantom{0} & \cellcolor[cmyk]{0.5,0,0,0}0 &\cellcolor[cmyk]{0.5,0,0,0} 1 & \cellcolor[cmyk]{0.5,0,0,0}1 & \cellcolor[cmyk]{0.5,0,0,0}1  \\ \cline{1-1} \cline{5-10}
\phantom{1} & \phantom{0} & \cellcolor[cmyk]{0,0,0.5,0}1 & \cellcolor[cmyk]{0,0,0.5,0}1 &\cellcolor[cmyk]{0.5,0.5,0.5,0} 1 & \cellcolor[cmyk]{0.5,0.5,0.5,0}1 & \cellcolor[cmyk]{0.5,0.5,0.5,0}0 & \cellcolor[cmyk]{0.5,0.5,0.5,0}1 & \cellcolor[cmyk]{0.5,0.5,0.5,0}0 &\cellcolor[cmyk]{0.5,0.5,0.5,0} 0 & \cellcolor[cmyk]{0.5,0,0,0}1 & \cellcolor[cmyk]{0.5,0,0,0}0 & \cellcolor[cmyk]{0.5,0,0,0}0 & \cellcolor[cmyk]{0.5,0,0,0}1  \\  \cline{4-4} \cline{10-10} \cline{14-14}
\phantom{1} & \phantom{1} & \cellcolor[cmyk]{0,0,0.5,0}0 & \phantom{0} &\cellcolor[cmyk]{0.5,0.5,0.5,0} 0 &\cellcolor[cmyk]{0.5,0.5,0.5,0} 1 &\cellcolor[cmyk]{0.5,0.5,0.5,0}1 & \cellcolor[cmyk]{0.5,0.5,0.5,0}1 & \cellcolor[cmyk]{0.5,0.5,0.5,0}0 & \phantom{1} & \cellcolor[cmyk]{0.5,0,0,0}1 & \cellcolor[cmyk]{0.5,0,0,0}0 & \cellcolor[cmyk]{0.5,0,0,0}1 & \phantom{0}  \\ \cline{1-4} \cline{9-9} \cline{13-14}
\cellcolor[cmyk]{0,0.5,0,0}0 & \cellcolor[cmyk]{0,0.5,0,0}1 & \cellcolor[cmyk]{0,0.5,0,0}0 & \cellcolor[cmyk]{0,0.5,0,0}0 & \cellcolor[cmyk]{0.5,0.5,0.5,0}1 &\cellcolor[cmyk]{0.5,0.5,0.5,0} 0 & \cellcolor[cmyk]{0.5,0.5,0.5,0}0 &\cellcolor[cmyk]{0.5,0.5,0.5,0} 1 & \phantom{1} & \phantom{0} & \cellcolor[cmyk]{0.5,0,0,0}1 & \cellcolor[cmyk]{0.5,0,0,0}1 & \cellcolor[cmyk]{0,0.5,0,0}1 & \cellcolor[cmyk]{0,0.5,0,0}1  \\ \cline{4-4} \cline{8-8} \cline{12-12}
\cellcolor[cmyk]{0,0.5,0,0}1 & \cellcolor[cmyk]{0,0.5,0,0}1 & \cellcolor[cmyk]{0,0.5,0,0}0 & \phantom{1} &\cellcolor[cmyk]{0.5,0.5,0.5,0} 1 &\cellcolor[cmyk]{0.5,0.5,0.5,0} 0 & \cellcolor[cmyk]{0.5,0.5,0.5,0}1 & \phantom{0} & \phantom{1} & \phantom{1} & \cellcolor[cmyk]{0.5,0,0,0}0 & \phantom{0} & \cellcolor[cmyk]{0,0.5,0,0}0 & \cellcolor[cmyk]{0,0.5,0,0}1  \\ \cline{3-3} \cline{7-12}
\cellcolor[cmyk]{0,0.5,0,0}0 & \cellcolor[cmyk]{0,0.5,0,0}1 & \phantom{1} & \phantom{0 }& \cellcolor[cmyk]{0.5,0.5,0.5,0}1 & \cellcolor[cmyk]{0.5,0.5,0.5,0}1 & \cellcolor[cmyk]{0,0.5,0.5,0}1 & \cellcolor[cmyk]{0,0.5,0.5,0}1 &\cellcolor[cmyk]{0,0.5,0.5,0} 0 &\cellcolor[cmyk]{0,0.5,0.5,0} 1 &\cellcolor[cmyk]{0,0.5,0.5,0} 0 & \cellcolor[cmyk]{0,0.5,0.5,0}0 & \cellcolor[cmyk]{0,0.5,0,0}1 & \cellcolor[cmyk]{0,0.5,0,0}0  \\ \cline{2-2} \cline{6-6} \cline{12-12}
\cellcolor[cmyk]{0,0.5,0,0}1 & \phantom{0} & \phantom{1} & \phantom{1} & \cellcolor[cmyk]{0.5,0.5,0.5,0}0 & \phantom{0} & \cellcolor[cmyk]{0,0.5,0.5,0}0 & \cellcolor[cmyk]{0,0.5,0.5,0}1 & \cellcolor[cmyk]{0,0.5,0.5,0}1 & \cellcolor[cmyk]{0,0.5,0.5,0}1 & \cellcolor[cmyk]{0,0.5,0.5,0}0 & \phantom{1} & \cellcolor[cmyk]{0,0.5,0,0}1 & \cellcolor[cmyk]{0,0.5,0,0}0  \\\hline
\end{array}$$
\end{scriptsize}
\end{table}

In general, we need to intercept
$$N=2^{L_1-1}+T-D=2^{L_1-1}(2^{L_2}-\ensuremath{\mathcal{Z}}_{\alpha}(1))$$
 bits of the shrunken sequence for the recovered triangles to overlap.
This number depends completely on the value of $\ensuremath{\mathcal{Z}}_{\alpha}(1)$, which depends on the primitive polynomial $p(x)$ (characteristic polynomial of the interleaved m-sequences).
This means that sometimes the number of needed intercepted bits will be greater than  suitable for practical applications.
For example, in Table~\ref{tab:6} we can see the different values of $\zech{\alpha}{1}$, for different primitive polynomials of degree $5$.
In order to make more difficult the recovery, we would use $p(x)=1+x+x^2+x^3+x^5$ since it produces the minimum value of $\zech{\alpha}{1}$.
In order to recover the sequence, we expect the cryptographer to use $q(x)=1+x^2+x^3+x^4+x^5$, which produces the maximum value of $\zech{\alpha}{1}$, and we have to intercept a smaller number bits to recover the complete sequence.

\begin{table}
\caption{Values of $\zech{\alpha}{1}$ for different primitive polynomials of degree $5$\label{tab:6}}
\[
\begin{array}{|c|c|}\hline
p_2(x)&\zech{\alpha}{1}\\ \hline \hline
x^5 + x^2 + 1 & 18\\\hline
x^5 + x^4 + x^2 + x + 1&19\\\hline
x^5 + x^3 + x^2 + x + 1 &12\\\hline
x^5 + x^3 + 1&14\\\hline
x^5 + x^4 + x^3 + x + 1&13\\\hline
x^5 + x^4 + x^3 + x^2 + 1&20\\\hline
\end{array}
\]
\end{table}

\subsection{Cryptanalysis}
In this section we introduce a cryptanalysis of the shrinking generator based on  the results given in Section~\ref{sec:mod}.
This attack is based on an exhaustive search over the initial states of the first register $R_1$.
Thus, the complexity of the brute-force attack trying all the possible keys is reduced by a factor $2^{L_2}$.

\subsubsection{General idea}

Given $n$ bits, $\pmb{s}=\{s_0, s_1, \ldots, s_{n-1}\}$, of the shrunken sequence, Algorithm~1
tests if a given initial state $\pmb{a}=\{a_0, a_1, \ldots, a_{L_1-1}\}$ for the register $R_1$ is considered correct or not.
The idea is to recover bits of the first interleaved m-sequence of the shrunken sequence.
If two different bits are stored in the same position, the initial state  $\pmb{a}$ is incorrect.
If $\pmb{a}$ is considered correct,  Algorithm~1 also returns the matrix $A$ with the value and the positions of the recovered bits in the first interleaved m-sequence.
Once we have recover a part of the first interleaved m-sequence, we can recover the complete shrunken sequence.


\begin{tabular}{l}
\hline\noalign{\smallskip}
\textbf{Algorithm 1.} \textbf{Crypto}: Test an initial state for $R_1$\\
\noalign{\smallskip} \hline \noalign{\smallskip}

\textbf{Input:}  $p_1(x)$, $p(x)$, $\delta$, $\pmb{s}$ and $\pmb{a}$\\

\medskip

\textbf{function} $[M, Stop]=$\textbf{SubCrypto}$(p_1(x),p(x), \delta, \pmb{s}, \pmb{a})$\\
01:\quad Compute  $\{a_i\}$ using $p_1(x)$ and $\pmb{a}$ until finding $n=length(\pmb{s})$ ones;\\
02:\quad Store in $P$ the positions of the $1$s in the generated bits of $\{a_i \}$;\\
03:\quad Store in $P$ the new positions computed as $P_i\cdot d \text{ mod } (2^{L_2}-1)$;\\
04:\quad Store $[P_i, s_i]$ in a matrix $M$;\\
05:\quad $Stop=1$;\\
06:\quad \textbf{while} $Stop=1$ and $n>1$\\
07:\quad \quad		 Update $P$ with the new positions;\\
08:\quad \quad		 Update $\pmb{s}$ with  $\{s_0+s_1, s_1+s_2, \ldots, s_{n-2}+s_{n-1}\}$;\\
09:\quad \quad		 Store [m,n]=size(M);\\
10:\quad \quad		 \textbf{for} $j=0$ to $m-1$\\
11:\quad \quad \quad      \textbf{for} $k=0$ to $length(P)-1$\\
12:\quad \quad \quad \quad    \textbf{if} $M_{j1}=P_k$\\
13:\quad \quad \quad \quad \quad \textbf{if} $M_{j2}\not = s_k$\\
14:\quad \quad \quad \quad \quad \quad   Initialise $M$;\\
15:\quad \quad \quad \quad \quad \quad   $Stop=0$;\\		
16:\quad \quad \quad \quad \quad \textbf{end if}\\
17:\quad \quad \quad \quad    \textbf{end if}\\
18:\quad \quad \quad \quad 	Store $[P_k, s_k]$ in $M$;\\
19:\quad \quad \quad      \textbf{end for} \\
20:\quad \quad		 \textbf{end for}\\
21:\quad \textbf{end while}\\
\textbf{end function}\\
\textbf{Output:}\\
  $M$: Recovered bits and their position in the first interleaved m-sequence.  \\
 $Stop$: $1$ if the initial state is considered correct and $0$ otherwise.\\
\noalign{\smallskip}\hline
\end{tabular}

\begin{remark}
It is worth pointing out that this work can be equivalently done with rule $60$, $x_{i}^{t+1}=x_{i-1}^{t}+x_{i}^{t}$.
In this case, the sequences would appear in reverse order along the CA, but the results would be identical.
\end{remark}

\subsubsection{Numerical example}
We are going to consider the advantageous case when $L_2=L_1+1$.
We know that $\delta=T_2-2$ (Corollary~\ref{cor:1}) and, furthermore,  $p(x)$ is the reciprocal polynomial of $p_2(x)$ (see~\cite{Cardell2016f}).
In this case, it is not necessary to compute $\delta $, which prevent us from doing more calculations.
We will also see that in this case, the computations of the logarithms on the first round are trivial.

\begin{example}\label{ex:4}
Consider two registers $R_1$ and $R_2$ with characteristic polynomials $p_1(x)=1+x+x^3+x^4+x^6$ and $p_2(x)=1+x^3+x^7$, respectively.

Assume we intercept $10$ bits of the shrunken sequence: $\pmb{s}=\{1, 0,  0,  0,  0,  1,  0,  0,  0,  0\}$.

Notice that, in this case, the period of the sequence  is $2^6(2^7-1)=8128$.

We apply Algorithm 1 in order to check if the initial state $a=\{1, 1, 1, 0, 1, 1\}$ is correct for $R_1$.

Since $L_2=L_1+1$, we know that $p(x)$ is the reciprocal polynomial of $p_2(x)$, that is, $p(x)=1+x^4+x^ 7$.

In this case, we have that the distance of decimation is $\delta=T_2-2=2^{7}-3=125$ (Corollary~\ref{cor:1}).

\textbf{Input}: $p_1(x)=1+x+x^3+x^4+x^6$, $p(x)=1+x^4+x^7$, $\delta=125$, $\pmb{a}=\{1\ 1\ 1\ 0\ 1\ 1\}$ and $\pmb{s}=\{1, 0,  0,  0,  0,  1,  0,  0, 0,  0 \}$.

We compute bits of the m-sequence generated by $R_1$ using $\pmb{a}$ and store the positions of the ones until we find $10$ ones:  $pos=\{0, 1, 2, 4, 5, 6, 8 ,10, 11, 13\}$.


The positions of the intercepted $10$ bits in the first  interleaved m-sequence are:
$$d^0_i=pos=\{0, 125, 123, 119, 117, 115, 111, 107, 105, 101\}$$.

Each bit of $\pmb{s}$ is stored in the respective position of $pos$. We store this information in the matrix $M$ and order by position:
\[
M^T=
\left[
\begin{array}{cccccccccc}
     0 & 101 & 105 & 107 & 111 & 115 & 117 & 119 & 123 & 125\\
     1 &   0 &   0 &   0 &   0 &   1 &   0 &   0 &   0 &  0\\
\end{array}
\right]
\]

We compute new positions and new bits to update the matrix $M$.
The new bits are computed applying rule 102 to the elements of $\pmb{s}$,
 $\{1\ 0\ 0\ 0\ 1\ 1\ 0\ 0 \ 0 \}$ and they are stored, respectively,  in the positions (mod 127):
\begin{align*}
d_0^1&=\ensuremath{\mathcal{Z}}_{\alpha}(125)=65\\
d_1^1&=\ensuremath{\mathcal{Z}}_{\alpha}(125-123)+123=2\zech{\alpha}{1}+123=63\\
d_2^1&=\ensuremath{\mathcal{Z}}_{\alpha}(123-119)+119=4\zech{\alpha}{1}+119=126\\
d_3^1&=\ensuremath{\mathcal{Z}}_{\alpha}(119-117)+117=2\zech{\alpha}{1}+117=57\\
d_4^1&=\ensuremath{\mathcal{Z}}_{\alpha}(117-115)+115=2\zech{\alpha}{1}+115=55\\
d_5^1&=\ensuremath{\mathcal{Z}}_{\alpha}(115-111)+111=4\zech{\alpha}{1}+111=118\\
d_6^1&=\ensuremath{\mathcal{Z}}_{\alpha}(111-107)+107=4\zech{\alpha}{1}+107=114\\
d_7^1&=\ensuremath{\mathcal{Z}}_{\alpha}(107-105)+105=2\zech{\alpha}{1}+105=45\\
d_8^1&=\ensuremath{\mathcal{Z}}_{\alpha}(105-101)+101=4\zech{\alpha}{1}+101=108\\
\end{align*}
Up to now, there are no repeated positions, then we continue with the next round.
The new bits to be stored are given by
$\{1, 0, 0,  1, 0, 1, 0, 0 \}$ and the next positions are:

\begin{align*}
d_0^2&=\ensuremath{\mathcal{Z}}_{\alpha}(123)=3\\
d_1^2&=\ensuremath{\mathcal{Z}}_{\alpha}(125-119)+119=2\zech{\alpha}{3}+119=111\\
d_2^2&=\ensuremath{\mathcal{Z}}_{\alpha}(123-117)+117=2\zech{\alpha}{3}+117=109\\
d_3^2&=\ensuremath{\mathcal{Z}}_{\alpha}(119-115)+115=4\zech{\alpha}{1}+115=122\\
d_4^2&=\ensuremath{\mathcal{Z}}_{\alpha}(117-111)+111=2\zech{\alpha}{3}+111=103\\
d_5^2&=\ensuremath{\mathcal{Z}}_{\alpha}(115-107)+107=8\zech{\alpha}{1}+107=121\\
d_6^2&=\ensuremath{\mathcal{Z}}_{\alpha}(111-105)+105=2\zech{\alpha}{3}+105=97\\
d_7^2&=\ensuremath{\mathcal{Z}}_{\alpha}(107-101)+101=2\zech{\alpha}{3}+101=93\\
\end{align*}

Position  111 appears again and we have to store 0 again. There is no contradiction.

In next round we have to store the bits
$\{1\ 0\ 1\  1\ 1\ 1\  0 \}$ in the positions $d_i^3$, $i=1, \ldots, 6$.
However, we know that it is easier to compute the positions
$d_i^4$, $i=1, \ldots, 5$ (see Remark~\ref{rem:2}) for the  new  bits $\{1\ 1\ 0\ 0\ 0\ 1\}$.

\begin{align*}
d_0^4&=\ensuremath{\mathcal{Z}}_{\alpha}(117)=90\\
d_1^4&=\ensuremath{\mathcal{Z}}_{\alpha}(125-115)+115=2\zech{\alpha}{5}+115=88\\
d_2^4&=\ensuremath{\mathcal{Z}}_{\alpha}(123-111)+111=4\zech{\alpha}{3}+111=95\\
d_3^4&=\ensuremath{\mathcal{Z}}_{\alpha}(119-107)+107=4\zech{\alpha}{3}+107=91\\
d_4^4&=\ensuremath{\mathcal{Z}}_{\alpha}(117-105)+105=4\zech{\alpha}{3}+115=89\\
d_5^4&=\ensuremath{\mathcal{Z}}_{\alpha}(115-101)+101=2\zech{\alpha}{7}+101=109\\
\end{align*}

Position 109 appears again and we have to store a 1, but there is another bit in this position with value 0. We have a contradiction, so the guessed initial state  is not correct.

Output: $STOP=0$
The initial state $\pmb{a}=\{1, 1, 0, 0, 0, 0, 0, 0\}$ is not correct since we found a contradiction.
\hfill $\blacksquare$
\end{example}

In this case, we have just needed to compute the logarithms  $\zech{\alpha}{1}$, $\zech{\alpha}{3}$, $\zech{\alpha}{5}$, $\zech{\alpha}{7}$, $\zech{\alpha}{117}$, $ \zech{\alpha}{123}$,  $\zech{\alpha}{125}$.
However, according to Section~\ref{sec:zech} we will see that we just need to compute two of these logarithms.

We start computing $\zech{\alpha}{1}=97$ and according to Remark~\ref{rem:1}, we have $\zech{\alpha}{97}=1$.
It is easy to see that $7$ and $97$ are in the same cyclotomic coset, since $4\cdot 97= 7 \text{ mod }127$.
Therefore, we have:
\[
\zech{\alpha}{7}=\zech{\alpha}{4 \cdot 7}= 4\zech{\alpha}{7}=4
\]

Now, we know that
$$\zech{\alpha}{123}-123=\zech{\alpha}{-123}=\zech{\alpha}{4}=4\zech{\alpha}{1}=7 \rightarrow \zech{\alpha}{123}=3$$
and then $\zech{\alpha}{3}=123$.
It is aso possible to check that 123 and 125 are in the same cyclic coset, since $125=2^6\cdot123\text{ mod } 127$.
Therefore, $\zech{\alpha}{125}=2^6\zech{\alpha}{123}=65$.

Finally, we need to compute $\zech{\alpha}{5}=50$.
\[
\zech{\alpha}{117}-117=\zech{\alpha}{-117}=\zech{\alpha}{10}=2\zech{\alpha}{5}= 100 \rightarrow \zech{\alpha}{117}=90.
\]
So computing $\zech{\alpha}{1}$ and $\zech{\alpha}{5}$ and due to the properties of Zech logarithm, we can deduce all the logarithms needed in this example.
%
%
%
%
In general, the number of logarithms we need to compute in order to obtain the contradiction we need is smaller than it seems, due to the properties showed in Section~\ref{sec:zech}.

\subsection{Recovering the initial state of $R_2$}


The idea is to use the recovered bits of the first interleaved m-sequence to recover the m-sequence produced by $R_2$.
According to Proposition~\ref{prop:1}, if we know this m-sequence, we can recover the initial state of the m-sequence $\{b_i\}$ generated by $R_2$.

\begin{example}
We consider Example~\ref{ex:4}.
Assume we apply Algorithm~1 with the correct initial state $\pmb{a}=\{1, 0, 0, 0, 0, 0\}$.
In this case, the algorithm returns STOP=1 and the matrix $M$ with the recovered bits.
In Appendix~\ref{ap:1}, we find the returned matrix $M$ and it is possible to see that we have recovered 46 bits and their positions in the first interleaved m-sequence of the shrunken sequence, which we denote now by $\{v_i\}$.
We know that these sequence has as characteristic polynomial $p(x)=1+x^4+x^7$ and the period is 127.
Therefore, we know that $v_{i}+v_{i+4}+v_{i+7}$, for $i\geq 0$ and then:
\begin{align*}
v_{46}+v_{50}+v_{53}=0 &\longrightarrow v_{46}=0\\
v_{47}+v_{51}+v_{54}=0& \longrightarrow v_{51}=1\\
\end{align*}
We known $12$ consecutive bits $\{v_{45}, v_{46}, \ldots, v_{56}\}$, of the first interleaved m-sequence $\{v_i\}$ and we know that  the characteristic polynomial of the sequence has degree $7$.
Thus, we compute the whole m-sequence $\{v_i\}$.
Finally according to Proposition~\ref{prop:1}, the initial state of $R_2$ will correspond to $\{b_0, b_{125}, b_{123}, b_{121}, , b_{119}, b_{117},$ $b_{115}\}$.\hfill $\blacksquare$
\end{example}

\subsection{Discussion}
In this algorithm, we have performed an exhaustive search over $2^{L_1-1}$ initial states of $R_1$, which reduces the complexity of a brute-force attack by a factor of $2^{L_2}$.
In Table~\ref{tab:30} some numerical results are depicted. 
We denote by $p_1(x)$ and $p_2(x)$  the characteristic polynomials
$R_1$ and $R_2$, respectively, $n$ is the number of intercepted bits, $T$ represents the period of the corresponding shrunken sequence and $N_{IS}$ denotes the number of $R_1$ initial states with no contradiction. From these results we can deduce that our algorithm presents two main advantages against other proposals.

First, compared with other probabilistic approaches (see, for example, \cite{Golic2001b}) the algorithm here presented is deterministic.
This means that depending on the number of intercepted bits, the set of the possible correct states can have different sizes, but the correct one is certainly contained in such a set.

Second, 
although the higher the degrees of the characteristic polynomials are, the more intercepted bits we need,  the required keystream length grows  linearly in the length of $R_2$ while the period of the shrunken sequence grows exponentially.
This means that the number  of intercepted bits $n$ needed for the attack  is very low compared with the period of the shrunken sequence.
This fact did not happen in other proposals like \cite{Johansson1998}.
Low requirement of intercepted bits is a quite realistic condition for practical cryptanalysis.

The number of  initial states for $R_1$ with no contradiction is very low compared with the number of initial states analysed, so it makes easier the checking of the true pair of initial states in both registers $R_1$ and~ $R_2$.

Finally, this algorithm is particularly adequate for parallelization. 
In fact, it is possible to divide the $2^{L_1-1}$ possible initial states into several groups and process each group of states separately.

The computation of Zech logarithms is the most time-consuming part of the algorithm.
However, we have seen in Section~\ref{sec:zech} that many of the properties of this discrete logarithm can be used to reduce the calculations. 
For instance, in Example~\ref{ex:4}, the algorithm had to compute ten logarithms, but due to the properties of the Zech logarithm, computing only two logarithms the problem  was solved.

Furthermore, at the end of Section~\ref{sec:com}, we saw that the computations of the positions $d_k^m$ can be computed  performing a simple addition $\mod (2^L_{2}-1)$.

\begin{table}[t]
\caption{Some numerical results for the algorithm \label{tab:30}}
\begin{scriptsize}
\begin{center}
\begin{tabular}{|c|c|c|c|c|}
  \hline
  $p_1(x)$ & $p_2(x)$ & $n$ & $T$ & $N_{IS}$ \\   \hline
  $1+x^2+x^3$ & $1+x^3+x^4$ & 8 & 60 & 1\\\hline
  $1+x^2+x^3$ & $1+x^3+x^5$ & 9 & 124 & 1\\\hline
  $1+x^2+x^5$         & $1+x+x^ 6$          & 11               & 1008 & 1\\\hline
  $1+x^3+x^5$ & $1+x+x^7$ & 13 & 2032 & 1\\\hline
  $1+x^2+x^5$         & $1+x^3+x^7$         & 14               & 2032&1 \\ \hline
  $1+x+x^6$ & $1+x^3+x^7$ & 16 & 4046 & 1\\\hline
  $1+x+x^7$           & $1+x^2+x^3+x^4+x^8$ & 16               & 16320&1 \\ \hline
  $1+x+x^7$           & $1+x^4+x^9$         & 16               & 32704&1\\ \hline
  $1+x^2+x^3+x^4+x^8$ &  $1+x^4+x^9$        & 17               & 65408 & 1\\ \hline
  $1+x^4+x^9$         & $1+x^3+x^{10}$      & 18               & 261888 &1 \\ \hline
  $1+x^4+x^9$         & $1+x^2+x^5+x^9+x^{10}$ & 19 & 261888 & 1\\\hline
  $1+x^2+x^{11}$      & $1+x+x^5+x^8+x^{12}$ & 27 & 4193280 & 3\\\hline
  $1+x^{9}+x^{10}+x^{12}+x^{13}$ & $1+x+x^2+x^5+x^6+x^{13}+x^{14}$ & 30 & 67104768 & 3\\\hline
  $1+x^{9}+x^{10}+x^{12}x^{13}$ & $1+x+x^4+x^{15}+x^{16}$ & 52 & 268431360 & 1\\\hline
  $1+x+x^2+x^{5}+x^{6}+x^{13}+x^{14}$ & $1+x^2+x^5+x^{14}+x^{15}$ & 40 & 268427264 & 126\\\hline
  $1+x^{2}+x^{5}+x^{14}+x^{15}$ & $1+x+x^4+x^{6}+x^{16}$ & 50 & 1073725440 & 29\\\hline
  $1+x+x^{4}+x^{15}+x^{16}$ & $1+x+x^2+x^6+x^{10}+x^{11}+x^{17}$ & 58 & 4294934528 & 206\\\hline
\end{tabular}
\end{center}
\end{scriptsize}
\end{table}

\section{Conclusions}\label{sec:con}

The shrinking generator was conceived and designed as a nonlinear keystream generator based
on maximum-length LFSRs. However, this generator can be modelled in terms of 102-CAs. The effort to introduce decimation in order to break the inherent linearity of the LFSRs
has been useless, since the shrunken sequence can be modelled as the output sequence of a
model based on linear CAs. In this work, we analyse a family of one-dimensional, linear, regular
and cyclic 102-CA that describe the behaviour of the shrinking generator. These CAs generate
a family of sequences with the same characteristic polynomial and the same period as the shrunken
sequence. Taking advantage of the linearity and the similarity between the sequences generated
by these CAs, we propose a cryptanalysis based on the exhaustive search among the initial states
of the first register.

A natural extension of this work is the generalization of this procedure to
other cryptographic sequence generators: (a) All the family of the decimation-based keystream generators (the self-shrinking
generator, the generalized self-shrinking generator or the modified self-shrinking generator\cite{Cardell2016c, Cardell2016d,Cardell2017b}). (b) The so-called interleaved
m-sequences, as they present very similar structural properties to those
of the sequences obtained from irregular decimation generators.

\appendix
\section{Tables}\label{ap:1}
\[
M=\left[
\begin{array}{cc}
     0 & 1\\
    12 & 1\\
    23 & 0\\
    24 & 1\\
    25 & 1\\
    32 & 0\\
    38 & 1\\
    40 & 0\\
    41 & 0\\
    43 & 1\\
    45 & 1\\
    47 & 0\\
    48 & 1\\
    49 & 0\\
    50 & 1\\
    52 & 1\\
    53 & 1\\
    54 & 1\\
    55 & 0\\
    56 & 1\\
    59 & 0\\
    60 & 1\\
    64 & 0\\
    66 & 0\\
    68 & 1\\
    78 & 1\\
    79 & 0\\
    87 & 0\\
    91 & 0\\
    95 & 0\\
    98 & 0\\
    99 & 1\\
   101 & 0\\
   103 & 0\\
   105 & 1\\
   107 & 0\\
   108 & 1\\
   109 & 0\\
   110 & 0\\
   111 & 0\\
   112 & 1\\
   114 & 0\\
   115 & 0\\
   117 & 0\\
   118 & 0\\
   119 & 0\\
   \end{array}
   \right]
   \]
\end{document}